\input harvmac
\noblackbox
\def\npb#1#2#3{{\it Nucl.\ Phys.} {\bf B#1} (19#2) #3}

\def\plb#1#2#3{{\it Phys.\ Lett.} {\bf B#1} (19#2) #3}
\def\prl#1#2#3{{\it Phys.\ Rev.\ Lett.} {\bf #1} (19#2) #3}

\def\prd#1#2#3{{\it Phys.\ Rev.} {\bf D#1} (19#2) #3}

\def\jmp#1#2#3{{\it J. Math.\ Phys.} {\bf #1} (19#2) #3}


\def\frac#1#2{{#1 \over #2}}

%

%

%
\Title{\vbox{\baselineskip12pt
\hbox{hep-th/0001145}
\hbox{CALT-68-2257}
\hbox{CITUSC/00-006}}}
{\vbox{\centerline{Probable Values of the Cosmological Constant}
\bigskip
\centerline{in a Holographic Theory}}}
\medskip\bigskip
\centerline{Petr Ho\v rava$^{1,2}$ and Djordje Minic$^{1,3}$}
\bigskip\medskip
\centerline{\it $^1$CIT-USC Center for Theoretical Physics}
\medskip
\centerline{\it $^2$California Institute of Technology, Pasadena, CA 91125, 
USA}
\centerline{\tt horava@theory.caltech.edu}
\medskip
\centerline{\it $^3$Department of Physics and Astronomy, University of 
Southern California}
\centerline{\it Los Angeles, CA 90089-0484, USA}
\centerline{\tt minic@physics.usc.edu}
\baselineskip18pt
\medskip\bigskip\medskip\bigskip\medskip
\noindent\baselineskip16pt
We point out that for a large class of universes, holography implies that the 
most probable value for the cosmological constant is zero.  In four 
spacetime dimensions, the probability distribution takes the Baum-Hawking 
form, $dP\sim\exp(cM_p^2/\Lambda)d\Lambda$.  
\Date{January 2000}
\lref\wein{S. Weinberg, {\it Rev.\ Mod.\ Phys.} {\bf 61} (1989) 1;\hfill\break
S.M. Carroll, W.H. Press and E.L. Turner, {\it Ann.\ Rev.\ Astron.\ and 
Astrophys.} {\bf 30} (1992) 499.}
\lref\data{A nice review of recent data is N.A. Bachall, J.P. Ostriker, S. 
Pelmutter and P.J. Steinhardt, {\it Science} {\bf 284} (1999) 1481.}
\lref\worm{S. Coleman, \npb{307}{88}{867}; \npb{310}{88}{643};
 S. Giddings and A. Stominger, \npb{306}{88}{890}; \npb{307}{88}{854}; 
T. Banks, \npb{309}{88}{493}; 
T. Banks, I. Klebanov and L. Susskind, \npb{317}{89}{665}.}
\lref\wormone{J. Polchinski, \plb{219}{89}{251};
V. Kaplunovsky, unpublished; W. Fischler and L. Susskind,
\plb{217}{89}{48}.}
\lref\wormtwo{W. Fischler, I. Klebanov, J. Polchinski and L. Susskind, 
\npb{327}{89}{157}.}
\lref\entro{S.W. Hawking, \prd{13}{76}{191};
J.D. Bekenstein, \prd{23}{81}{287}.}
\lref\holo{G. 't Hooft, gr-qc/9310026; L. Susskind, \jmp{36}{95}{6377}, 
hep-th/9409089.}
\lref\banks{T. Banks, hep-th/9601151.}
\lref\hft{P. Ho\v rava, \prd{59}{99}{046004}, hep-th/9712130.}
\lref\wash{A.G. Cohen, D.B. Kaplan and A.E. Nelson, \prl{82}{99}{4971}, 
hep-ph/9803132.}
\lref\ver{C.P. Burgess, R.C. Myers and F. Quevedo, hep-th/9911164; H.
Verlinde and E. Verlinde, hep-th/9912018; C. Schmidhuber, hep-th/9912156.}
\lref\bousso{R. Bousso, hep-th/9905177; hep-th/9906022; hep-th/9911002.}
\lref\landlif{L.D. Landau and E.M. Lifshitz, {\it Statistical Physics},
part 1 (Pergamon, 1988).}
\lref\baumhaw{E. Baum, \plb{133}{83}{185}; S. W. Hawking, \plb{134}{84}{403}.}
\lref\linde{A.D. Linde, \plb{175}{86}{395}, \plb{202}{88}{194}.}
\lref\gh{G.W. Gibbons and S.W. Hawking, \prd{15}{77}{2738}.}
\lref\bryo{J.D. Brown and J.W. York, \prd{47}{93}{1420}, gr-qc/9209014.}
\newsec{Introduction}

One of the central problems of theoretical physics is why the cosmological 
constant is small \wein.

The cosmological constant problem has a twofold meaning:  it is a problem of 
fundamental physics, because the value of the cosmological constant $\Lambda$ 
is tied to vacuum energy density.  On the other hand, the cosmological 
constant tells 
us something about the large scale behaviour of the universe, since a small 
cosmological constant implies that the observable universe is big and (nearly) 
flat.  The problem is that there is an enormous discrepancy between the value 
of the vacuum energy density as predicted by quantum field theory of the 
standard-model degrees of freedom, and the cosmologically observed value of 
$\Lambda$ \data .  This discrepancy occurs already at very low energy scales,  
of order eV, and clearly represents the most flagrant naturalness problem in 
today's physics.    

Thus, the cosmological constant relates the properties of the microscopic 
physics of the vacuum to the long-distance physics on cosmic scales.%
\foot{This general philosophy was stressed in the wormhole approach to the 
cosmological constant problem (see \worm\ for the original references and  
\refs{\wormone,\wormtwo} for a critique of this approach).}
Therefore, the observable smallness of the cosmological constant should tell 
us something fundamental about the underlying microscopic theory of nature.

In this note we study implications of holography \refs{\holo,\entro}, taken 
as a fundamental property of the microscopic theory of quantum gravity, for 
the cosmological constant problem.  Assuming that the cosmological constant 
is a dynamical variable, and that holographic entropy can be given a 
Botzmannian interpretation, we point out that the most probable value of the 
cosmological constant in a holographic theory is zero, in ensembles of 
universes with finite-area holographic screens.  

The argument is very simple, but apparently has not been presented in the 
literature before.

\newsec{Holography and the Cosmological Constant}

It has been suggested on rather general grounds \refs{\banks,\hft,\wash} 
that holography should be relevant for solving the cosmological constant 
problem.  In its simplest heuristic form, this argument can be stated as 
follows.  The cosmological constant problem in local quantum field theory 
is a naturalness problem, following from a gross overcount of the degrees of 
freedom of the vacuum.  In a holographic theory, the Bekenstein-Hawking bound  
\entro\ imposes a natural limit on the number of degrees of freedom, which 
subsequently reduces the vacuum energy density in the microscopic theory.  
Notice that this intuitive argument does not suggest that $\Lambda$ should be 
zero; instead, it would make a small but non-zero cosmological constant 
natural.%
\foot{Various other aspects of the cosmological constant problem in the light 
of holography have been recently studied in \ver.}

In order to make this argument work on a practical level, we may need a 
precise microscopic model that exhibits holography in a manifest way.  
A phenomenological model of holography has been proposed in \hft , in the 
context of M-theory.  In that model one is lead, schematically, to the 
following expansion of the low-energy effective action in four space-time 
dimensions%
\foot{The phenomenological theory in \hft\ is formulated in eleven space-time 
dimensions; here we have implicitly compactified the theory to four dimensions 
on $T^7$.}
\eqn\eezero{S\approx NM^2\int d^4x\sqrt{g}\left(M^2+R+\frac{1}{M^2}R^2+
\ldots\right)}
(in this expansion we ignore various multiplicative constants of order one).  
Here $M$ is an {\it infrared\/} scale, essentially the inverse size of the 
system; $N$ is tied to the number of degrees of freedom in the theory;  
and $R^2$ symbolically denotes the terms quadratic in the Riemann tensor.  
The low-energy (super)gravity regime is defined as the regime in which the 
Einstein-Hilbert term is kept finite; this determines the Newton constant 
$G_N$ in terms of the infrared mass $M$ and the number of degrees of freedom 
$N$ 
\eqn\eezeroandhalf{G_N{}^{-1}\approx NM^2.}
This can be interpreted as the Bekenstein-Hawking formula for $N$ in terms of 
the area of the surface surrounding the system, measured in Planck units.  
Another indication for holography comes from the following observation.  
The Einstein-Hilbert term is dominant only if the higher-order curvature 
terms are surpressed, which happens for $R\ll M^2$.  This bound coincides with 
the Bekenstein bound \hft .  For exactly the same reason the cosmological 
constant term in \eezero\ is naturally small, of order 
$\Lambda\approx M^2$ -- the ``Hubble radius'' of the system.  

Thus, the phenomenological approach of \hft\ seems to suggest at a somewhat 
semi-quantitative level that the cosmological constant problem -- as a 
naturalness problem -- could indeed be solved in a holographic theory (for 
other arguments from a different point of view, see \banks ).  However, the 
problem of finding a truly microscopic theory that manifestly satisfies 
holography still remains a fascinating challenge that has not been 
satisfactorily met so far.  

Short of a microscopic formulation of a holographic theory, we choose 
a different strategy to address the cosmological constant problem.  We 
simply {\it assume\/} that holography is valid, and show that in combination 
with certain robust thermodynamic arguments, holography indeed implies that 
{\it the most probable value\/} of the cosmological constant is zero.  

For concreteness we work in four space-time dimensions, but the argument can 
be easily generalized to other cases as well.  

At large scales gravity is described by the most general local effective 
action which incorporates the four-dimensional diffeomorphism invariance 
(the Einstein-Hilbert action, the cosmological constant term, plus higher 
order terms in the curvature tensor and its derivatives)
\eqn\eeone{S_{\rm eff}=-\int d^4x\sqrt{g}\left({1 \over {8 \pi G}}\Lambda+ 
{1 \over {16 \pi G}}R+aR_{\mu\nu}R^{\mu\nu}+bR^2+cR_{\mu\nu\rho\sigma}
R^{\mu\nu\rho\sigma}+\ldots\right).}
Note that $\Lambda$ denotes an effective cosmological constant, which 
also takes into account the vacuum energy density of matter.  We also assume 
that $\Lambda$ is a dynamical variable, without specifying a particular 
mechanism that leads to a dynamical $\Lambda$.  Several such mechanisms 
are available in the literature \wein : coupling gravity to a three-form 
gauge field \baumhaw , topology change \worm, and chaotic inflation \linde ,  
to name a few.  

According to the holographic principle \refs{\holo,\bousso}, the entire 
information about the space-time can be stored on particular hypersurfaces 
(called holographic screens) \bousso , such that the total number of 
degrees of freedom living on these holographic screens does not exceed the
Bekenstein-Hawking bound.  At present this is the only available formulation 
of the holographic principle, but it will be sufficient for our argument.  

Let us concentrate on a class of universes with well-defined  holographic 
screens of finite area, examples of which were explicitly constructed for 
various space-time geometries in \bousso.  In this class of universes, 
a closer look at Einstein's equations leads to the following scaling relation 
between the characteristic size $r$ of the preferred screen and the 
cosmological constant 
\eqn\eefour{\Lambda r^2 \approx 1.}
As can be seen in \bousso , this relation between $\Lambda$ and $r$ is valid 
for a broad class of holographic screens in spacetimes with non-negative 
$\Lambda$, including the de~Sitter spacetime, the Einstein static universe, 
and a class of the Robertson-Walker cosmologies.  In the de~Sitter case, 
for example, the preferred screen is the cosmological horizon \bousso , whose 
area is given by \gh\ 
\eqn\eedesitter{A=\frac{12\pi}{\Lambda}.}
In the anti-de~Sitter spacetime, the size of the holographic screen is 
infinite, and we do not know whether our argument can be extended to the cases 
with holographic screens of infinite area.   

The holographic principle asserts that the total number of degrees of freedom, 
or entropy $S$, living on the holographic screen is bounded by one quarter of 
the area in Planck units, 
\eqn\eefive{S\approx r^2 M_p^2.}
Of course, this formula is just an upper bound, and should be really treated 
as an inequality. We will take this into account below.

Using the scaling relation \eefour\ valid for the holographic screens, 
together with the Bekenstein-Hawking bound \eefive , we get an intriguing 
formula which relates the cosmological constant and the holographic 
gravitational entropy $S$
\eqn\eesix{\Lambda S \approx M_p^2.}
More precisely, this formula should be viewed as an inequality
\eqn\eeseven{S \leq {M_p^2 \over \Lambda}.}
Thus, the holographic gravitational entropy (or the total number of degrees of 
freedom living on the holographic screens) $S$ increases with a decreasing 
cosmological constant.  

We use this observation to argue that the most probable value for the 
cosmological constant is zero. The argument proceeds as follows.  

At large distances the cosmological constant could be treated as a classical 
variable (its fluctuations can be neglected). Thus, following the 
Boltzmann principle from statistical mechanics \landlif, the probabity 
distribution for the cosmological constant $\Lambda$ to have a value in the 
interval from $\Lambda$ to $\Lambda + d\Lambda$ is given by
\eqn\eeeight{w({\Lambda}) d\Lambda={\it const}\,\exp\left\{S(\Lambda)\right\}}
(with the Boltzmann constant set to one).  In \eeeight ,  $S(\Lambda)$ is the 
holographic entropy, which can be formally regarded as a function of 
$\Lambda$ using \eeseven .  We conclude that 
\eqn\eefellini{w({\Lambda})d\Lambda={\it const}\,\exp\left\{{cM_p^2\over
\Lambda}\right\}.}
Here $c$ denotes a constant of order one which takes into account the 
neglected numerical factors.  This formula tells us that the probability 
distribution is strongly peaked around the value $\Lambda =0+$.  Hence, the 
most probable value for the cosmological constant, as implied by holography 
and thermodynamics, is zero.  

We have obtained this formula by saturating the Bekenstein-Hawking bound,   
but it is easy to see that even if the holographic
bound is taken as a true inequality the same conclusion follows, due to 
the properties of the function $\exp\{{cM_p^2 \over \Lambda}\}$, and the fact 
that the probability distributions have to be normalized to one.  Notice also 
that we have implicitly worked with a microcanonical ensemble of universes, 
given our Boltzmannian interpretation of the holographic entropy.%
\foot{Some aspects of the microcanonical ensemble for gravity have been 
discussed in \bryo .}

\newsec{Discussion}

Here we compare our result with the Baum-Hawking mechanism for the 
vanishing of the cosmological constant.

Our probability distribution \eefellini\ exhibits the same exponential 
dependence on $1/\Lambda$ as the probability formula obtained in Euclidean 
quantum gravity by Baum and Hawking \baumhaw\ .%
\foot{In $D$ spacetime dimensions, our probability distribution generalizes to 
$\exp\left\{\tilde c M_p^{D-2}\Lambda^{(2-D)/2}\right\}$, and therefore gives 
a functional dependence on $\Lambda$ which again agrees with the result of the 
Euclidean path integral approach.}
In the case of de~Sitter universes, the exact value of the numerical constant 
$c$ in \eefellini\ follows from \eedesitter , and is found to be $c=3\pi$.  
Thus, for de~Sitter universes, our holographic probability distribution 
exactly coincides with the Baum-Hawking distribution.

Recall that the Baum-Hawking mechanism asumes the validity of the Euclidean 
effective action formalism in quantum gravity; the Minkowski action would lead 
to an oscillating factor $\exp[i {M_p^2 \over \Lambda}]$ which completely 
changes the conclusion about the most probable value for the cosmological 
constant. Also, the Euclidean action for Einstein's gravity is well known to 
be unbounded from below; this fact renders the Baum-Hawking  mechanism rather 
problematic.  Moreover, it is not clear -- at least within the Hamiltonian 
approach as presented in \wormtwo\ -- whether the Baum-Hawking factor 
$\exp[{c M_p^2 \over\Lambda}]$ can be interpreted as a probability 
distribution in the context of Euclidean quantum gravity with spacetime 
topology change.  

By contrast, our argument is based on the assumptions of holography (without 
specifying its microscopic origin) and thermodynamics.  According to the 
holographic principle, the holographic entropy counts the total number of 
degrees of freedom in quantum gravity. The use of the Boltzmann formula is 
therefore justified by the microscopic definition of entropy. Likewise, the 
use of the problematic Euclidean formalism of quantum gravity has been 
completely avoided by the use of the holographic principle.  Nevertheless, we 
find it intriguing that the same probability distribution is found in both 
cases.  

In this paper, we have presented a simple argument suggesting that in a 
holographic theory, the cosmological constant can be naturally small.  
This conclusion follows from the simple but somewhat subtle fact that by 
maximizing the entropy in a holographic theory, one minimizes the vacuum 
energy density, which indeed seems contrary to the intuitive expectation 
based on our experience with local field theory, where large entropy 
gives a large contribution to the vacuum energy density.

\bigskip\medskip\noindent
We wish to thank T. Banks, J. Polchinski, A. Strominger, and E. Witten for 
useful discussions.  
The work of P.H. has been supported in part by a Sherman Fairchild Prize 
Fellowship, and by DOE grant DE-FG03-92-ER~40701.  The work of D.M. has been 
supported in part by the US Department of Energy under grant number 
DE-FG03-84ER~40168.  

\listrefs
\end